\documentclass[12pt,onecolumn]{IEEEtran}
\usepackage{graphicx}
\usepackage[caption=false,font=footnotesize]{subfig}
\usepackage{amsfonts}
\usepackage{algorithm}
\usepackage{algorithmic}

\newcommand{\mbf}[0]{\mathbf{f}}
\newcommand{\mbg}[0]{\mathbf{g}}
\newcommand{\mbh}[0]{\mathbf{h}}

\begin{document}

\title{\textbf{Experimental Performance Evaluation of Location Distinction for
        MIMO Channels}}

\author{\IEEEauthorblockN{Dustin Maas\IEEEauthorrefmark{1},
Neal Patwari\IEEEauthorrefmark{1},
Daryl Wasden\IEEEauthorrefmark{1}, \\
Sneha K. Kasera\IEEEauthorrefmark{2} and
Michael A. Jensen\IEEEauthorrefmark{3}}\\
\IEEEauthorblockA{\IEEEauthorrefmark{1}Dept. of Electrical and
    Computer~Engineering   }
\IEEEauthorblockA{\IEEEauthorrefmark{2}School of Computing}\\
\IEEEauthorblockA{University of Utah, Salt Lake City, USA}\\
\IEEEauthorblockA{[maas@ece,npatwari@ece,wasden@ece,kasera@cs].utah.edu}\\
\IEEEauthorblockA{\IEEEauthorrefmark{3}Dept. of Electrical and Computer~Engineering
Brigham Young University, Provo, Utah, USA\\
jensen@ee.byu.edu}}

\maketitle
\thispagestyle{empty}

\begin{abstract} 
Location distinction is defined as determining whether or not the position of a
device has changed. We introduce methods and metrics for performing location
distinction in multiple-input multiple-output (MIMO) wireless networks. Using
MIMO channel measurements from two different testbeds, we evaluate the
performance of temporal signature-based location distinction with varying
system parameters, and show that it can be applied to MIMO channels with
favorable results. In particular, a 2x2 MIMO channel with a bandwidth of 80 MHz
allows a 64-fold reduction in miss rate over the SISO channel for a fixed false
alarm rate, achieving as small as $4\times 10^{-4}$ probability of false alarm
for a $2.4\times 10^{-4}$ probability of missed detection. The very high
reliability of MIMO location distinction enables location distinction systems
to detect the change in position of a transmitter even when using a single
receiver.

\end{abstract} 

\section{Introduction}
\label{sec:introduction}
Location distinction is defined as determining whether or not the position of a
device has changed. In the context of a wireless network, this means detecting
when a transmitter changes its position via measurements made at one or more
receivers, or vice versa. 

Location distinction is fundamentally different from localization, in that
location distinction is not concerned with the position of the transmitter,
only whether or not it has moved. Location distinction should work under two
use cases: (1.) when a wireless device is continuously moving; and (2.) when a
wireless device and access point are stationary for a long time and suddenly a
transmission with the same claimed identity is sent from a new location.  Under
use case (1.) the algorithm should detect a new location with each
transmission, while under use case (2.) the algorithm should decide the new
transmission is from a different location. 
 
The ability to perform location distinction provides several benefits,
including an improved capability to monitor the positions of radio-tagged
objects, better energy conservation in radio localization systems, and a means
to detect impersonation attacks in wireless networks
\cite{patwari2007robust,sheng2008detecting}. Location distinction has been
shown to be useful in detecting the Sybil attack
\cite{xiao2009channel,demirbas2006rssi}. Other work has also shown that
characteristics of the physical layer of wireless networks, such as received
signal strength (RSS), channel impulse response (CIR), or channel frequency
response can be exploited to detect changes in transmitter/receiver positions
\cite{faria2006detecting, li2006securing, zhang2008advancing, xiao2008mimo}.  

Multiple-input multiple-output (MIMO)-capable devices represent the
state-of-the-art in wireless networking and have enabled significantly improved
spectral efficiencies in wireless networks. Many new wireless standards, such
as 802.11n, WiMax, and 4G cellular, take advantage of MIMO technology.
Enhancing these standards with the capability to perform location distinction
would offer extra security against impersonation attacks. For example, the
802.11n standard is vulnerable to impersonation and denial-of-service attacks
because the MAC addresses of network clients are sent over the air unencrypted
and may be eavesdropped on and used by an attacker in order to masquerade as a
legitimate client. 

Previous work has suggested using channel measurements gathered between a single
transmitter and multiple receivers in order to perform location distinction
\cite{faria2006detecting,patwari2007robust,zhang2008advancing,demirbas2006rssi,chen2007detecting}.
However, in typical WiFi networks, adjacent access points are set to operate on
different channels in order to reduce interference and a client operates on a
single channel. This makes collecting channel data at multiple access points
difficult. Extending location distinction to MIMO allows robust location
distinction to be performed with a single receiver.

\begin{figure}[htbp]
\centering
\includegraphics[width=3.3in ]{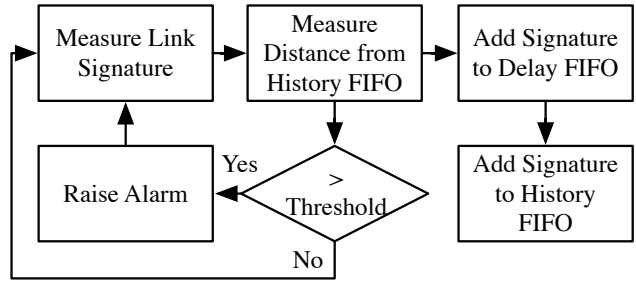}
\caption{Location distinction measures link signatures from received packets,
    and then raises an alarm if the current measurement differs greatly from
    those in the history.}
\label{F:LocDisDiagram}
\end{figure}

This paper evaluates the performance of the general location distinction
algorithm shown in Figure \ref{F:LocDisDiagram}, in which channel impulse
response measurements, called link signatures, are measured over time for a
given link, and each new link signature is compared to those in a history of
previous measurements in order to detect changes in position.  To the authors'
knowledge no implementation and experimental evaluation of MIMO-based location
distinction has been performed. We present the following work in order to
characterize the performance of temporal signature-based location distinction
in the context of a MIMO channel:
\begin{enumerate}
    \item We introduce MIMO temporal link signatures for quantifying the state
        of the MIMO channel.
    \item We perform two measurement experiments with two different
        experimental testbeds in order to evaluate location distinction under
        two distinct use cases.
    \item We evaluate spatially dense channel measurements in order to study the
        spatial evolution of temporal link signatures.
    \item We evaluate several trade-offs between system design parameters and
        performance, including: link signature history size, bandwidth, complex
        vs. magnitude-only signatures, use of delay between measurements, and number of
        antenna elements.
\end{enumerate}

The results show that MIMO location distinction algorithms
perform well in a variety of experimental conditions. For example, we
achieve a $4\times 10^{-4}$ probability of false alarm for a $2.4\times 10^{-4}$
probability of missed detection using a 2x2 MIMO channel with a bandwidth of 80
MHz, and a $3\times 10^{-4}$ probability of missed detection for a false alarm
rate of 0.01 using a 1x2 SIMO channel with a bandwidth of 20 MHz.

Additionally, we show that:
\begin{enumerate}
    \item In the context of spatially dense link signature measurements
        (inter-measurement distances $< \lambda$), it is necessary to introduce
        a delay between past and current measurements in order to reliably
        perform location distinction. The size of delay depends on the spatial
        density of the measurements.
    \item The number of link signatures to store in the history depends on
        the amount of temporal variation in the link signatures when the
        wireless device is stationary. 
    \item The most significant performance gain for MIMO
        vs. SISO location distinction occurs in the move from SISO to 2x2 MIMO.
        Further increasing the number of antenna elements offers diminishing
        returns. 
    \item When random phase shifts due to imperfect synchronization are
        removed, complex link signatures lead to better performance than magnitude-only
        link signatures.
    \item Increasing the bandwidth of the link signatures offers diminishing
        returns after about 20 MHz. In fact, higher bandwidth measurements are
        more susceptible to synchronization errors.
\end{enumerate}

This paper is organized as follows. In Section~\ref{sec:methods}, we describe
the link signatures, metrics, and MIMO location distinction algorithm. In
Section \ref{sec:Measurements}, we discuss two measurement experiments, which
we will refer to as Experiment I and Experiment II. In
Section~\ref{sec:results}, we present testing results and analysis of the MIMO
location distinction algorithm. We discuss related work in in
Section~\ref{sec:related_work}.  Conclusions and future work are presented in
Section~\ref{sec:conclusion}.

\section{\textbf{Methods}} 
\label{sec:methods} 
In this section, we first describe the link signatures we use for location
distinction and the difference metrics we use to quantify changes in them. Next,
we present a real-time location distinction algorithm and the framework 
for testing this algorithm.

\subsection{Link Signatures} 
\label{sub:link_signatures}
We define the $n$th \emph{complex temporal link signature} (CTLS) calculated for
the $c$th transmitter/receiver antenna
pair as
\begin{equation} \label{E:cplxTmpLinkSig}
	\mbf^{(n)}_{c} = [h_{c}^{(n)}(0),h_{c}^{(n)}(1 T_{s}),
    \ldots,h_{c}^{(n)}((M-1) T_{s})]
\end{equation}
where $h_c^{(n)}(\tau)$ is the band-limited channel impulse response as a function of
delay $\tau$, $M$ is the number of samples, $T_s$ is the
sampling period, and $c \in S$, where 
\begin{equation}
    S = \{1,...,k_1\}\times\{1,...,k_2\}.
\end{equation}

The number of transmitter and receiver antennas are represented by $k_1$ and
$k_2$, respectively. We also define the $n$th \emph{temporal link signature}
(TLS) calculated for the $c$th transmitter/receiver antenna pair as
\begin{equation} \label{E:tmpLinkSig}
	\mbg^{(n)}_{c} = [|h_{c}^{(n)}(0)|,|h_{c}^{(n)}(1 T_{s})| 
    \ldots,|h_{c}^{(n)}((M-1) T_{s})|].
\end{equation}

The MIMO channel measurements used in this paper are gathered using either a
multitone probe or preamble-based channel estimation, both of which are
described in Section \ref{sec:Measurements}. In both cases, time-domain
representations of the channel response are used for link signatures. 

We let the $n$th \emph{MIMO complex temporal link signature} (MIMO CTLS)
be the concatenation of the set of complex temporal link signatures measured
between the first $k_1 \times k_2$ transmitter and receiver antennas: 
\begin{equation}\label{E:cplxMIMOtmpLinkSig}
	\mathbf{F}^{n} = [\mbf^{(n)}_{c_1},\ldots,\mbf^{(n)}_{c_k}],
\end{equation}
where $c_1,...,c_k$ is a list of the elements of $S$.

Finally, we let the $n$th \emph{MIMO temporal link signature} (MIMO TLS) be the
concatenation of the set of temporal link signatures measured between the first
$k_1 \times k_2$ transmitter and receiver antennas: 

\begin{equation}\label{E:MIMOtmpLinkSig}
	\mathbf{G}^{n} = [\mbg^{(n)}_{c_1},\ldots,\mbg^{(n)}_{c_k}].
\end{equation}


\subsection{Difference Metric} 
\label{sub:metric}
In this section, we define the metric for measuring the difference between the current MIMO link
signature the FIFO history of previous MIMO link signatures below. The FIFO
history $\mathcal{H}$ for the previous $N$ MIMO link signatures is defined as 
\begin{equation} \label{E:cplxHist}
	\mathcal{H} = \{\mathbf{F}^{n}\}_{n=1}^{N}
\end{equation}
or
\begin{equation} \label{E:Hist}
	\mathcal{H} = \{\mathbf{G}^{n}\}_{n=1}^{N}
\end{equation}
depending on the MIMO link signature being used. The difference metric we
explore in this paper is
\begin{equation} \label{E:l2Distance}
    \Delta(\mathbf{F}^{N+D}, \mathcal{H}) = \frac{1}{\sigma} 
    \min_{\mathbf{F} \in \mathcal{H}} \| \mathbf{F} - 
    \mathbf{F}^{N+D} \|
    \label{E:distMetric}
\end{equation}
where $\sigma$ is the average distance between link signatures in the history, defined as
\begin{equation} \label{E:tprlDistance}
  	\sigma = \frac{1}{(N-1)(N-2)}\sum_{\mathbf{F}^m,
        \mathbf{F}^n \in 
    \mathcal{H}} \| \mathbf{F}^m - \mathbf{F}^n \|
    \label{E:distNorm}
\end{equation}
and $D$ is a delay parameter. This delay is inserted to increase the time between
the current link signature measurement and those in the history. As we show in
Section \ref{sub:physical}, $D > 1$ helps detection performance under use
case (1.). In the case of the magnitude-only TLS, the norms in
(\ref{E:distMetric}) and (\ref{E:distNorm}) are the $\ell_2$ norm; for the CTLS,
these norms are the $\phi_2$ norm, defined as 
\begin{equation} \label{E:phi2} 
    ||\mbg - \mbh||_{\phi_2} = \min_\phi ||\mbg - \mbh e^{j \phi}||_{\ell_2} =
    ||\mbg\|^2+\|\mbh\|^2 - 2\|\mbg^*\mbh\|.  
\end{equation} 
The $\phi_2$ norm removes the effect of random phase shifts that occur between
subsequent CTLS measurements.

We examine various sizes for the FIFO history $\mathcal{H}$ and the delay $D$
in Sections \ref{sub:history} and \ref{sub:physical} respectively. Changing
these parameters dramatically affects the detection performance of the location
distinction algorithm.  The delay has the effect of increasing the difference
between the latest link signature and those stored in the history. This is
beneficial for location distinction under use case (1.). The FIFO
history size is chosen to maximize the probability of detecting a change in
receiver position, while minimizing the probability of misidentifying a
stationary receiver as moving.

\subsection{Real-time Location Distinction} 
\label{sub:real_time}
A real-time location distinction algorithm is defined by the following steps:
\begin{enumerate}
	\item Measure the current link signature.
	\item Calculate the minimum difference $\Delta$ between the current link signature
        and the link signatures in the FIFO history $\mathcal{H}$.
	\item Compare the minimum difference $\Delta$ to a threshold $\gamma$. If 
        $\Delta > \gamma$, raise an alarm to indicate that the receiver has moved
        since the last link signature was measured. If $\Delta < \gamma$, do not
        raise an alarm, thereby indicating that the receiver has not moved 
        since the last link signature was measured.
	\item Add the current link signature to a FIFO delay buffer and add the 
        oldest link signature in the delay buffer to the FIFO history 
        $\mathcal{H}$.
	\item Return to step 1.  
\end{enumerate}

The process is illustrated in Figure \ref{F:LocDisDiagram}. This is a real-time
algorithm, but we note that in this paper, we first collect all of the link
signatures, and then evaluate location distinction in post-processing.

\subsection{Performance Evaluation} 
\label{sub:framework}
In this Section, we construct a framework used to apply the metrics described
in Section \ref{sub:metric} to the link signatures described in Section
\ref{sub:link_signatures} in order to test the performance of MIMO location
distinction. The performance evaluation is conducted using the following steps:
\begin{enumerate}
    \item The output of the difference metrics $$E_{f}^{N+D} =
        \Delta(\mathbf{F}^{N+D}, \mathcal{H})$$ and $$E_{g}^{N+D} =
        \Delta(\mathbf{G}^{N+D}, \mathcal{H})$$ are recorded for  
        stationary and moving receivers.
    \item We identify the probability of false alarm $P_{FA}$ and probability
        of detection $P_D$ for each antenna subset in reference to a possible
        difference threshold $\gamma$. We define the null and alternate
        hypotheses, $\mathbb{H}_0$ and $\mathbb{H}_1$ as follows:
	\begin{center}
		\begin{description}
			\item[$\mathbb{H}_0:$] Receiver has not moved.
  			\item[$\mathbb{H}_1:$] Receiver has moved. 
		\end{description}
	\end{center}
We treat $E_{f}^{N+D}$ and $E_{g}^{N+D}$ as random variables and denote their
conditional density functions under the two events above as $f_{E}(x |
\mathbb{H}_0)$ and $f_{E}(x | \mathbb{H}_1)$. The $E_{f}^{N+D}$ and
$E_{g}^{N+D}$ for a stationary and moving receiver are used to characterize
$f_{E}(x | \mathbb{H}_0)$ and $f_{E}(x | \mathbb{H}_1)$ respectively. We
calculate $P_{FA}$, $P_D$, and $P_M$ as:
	\begin{eqnarray}
		P_{FA} &=& \int_{x=\gamma}^\infty f_{E}(x | \mathbb{H}_0) dx  \nonumber \\
		P_{D}  &=& \int_{x=\gamma}^\infty f_{E}(x | \mathbb{H}_1) dx  \nonumber \\
        P_{M}  &=& 1-P_{D} \nonumber
	\label{E:PFA_PD}
	\end{eqnarray}
\end{enumerate}

The $P_{FA}$ and $P_D$ as a function of $\gamma$ allow us to evaluate how well
location distinction would have worked if a threshold of $\gamma$ was used in
the real-time algorithm. Thus the set of possible $P_{FA}$/$P_D$ combinations
provide a curve of feasible real-time detection performance.

\section{Measurements} 
\label{sec:Measurements}
We describe two MIMO measurement experiments. One is performed at Brigham Young
University \cite{wallace2006time}, and another is performed at the University
of Utah.  These datasets are used to evaluate the location distinction
algorithm according to the framework described in the previous section.

These experiments provide an opportunity to examine the following two use cases
for location distinction:
\begin{enumerate}
    \item A wireless device sends packets while in motion so that each new
        packet is sent from a distinct location. In this case, the location
        distinction algorithm should detect the change with every new packet.
        Our Experiment I provides MIMO data to test the performance of location
        distinction in this use case. In fact, measurements are made with fine
        enough spacial resolution that it is necessary to delay inserting the
        most recent measurements into the history FIFO in order to ensure
        sufficient decorrelation between a current measurement and those in the
        history.
    \item A wireless device sends packets while stationary for a long period of
        time. Then, a new packet is sent from a distinct location, either
        because the wireless device has moved, or because a second wireless
        device is attempting to impersonate the first from a different
        location. In either case, the location distinction algorithm should
        detect the change. Our Experiment II provides MIMO data to test the
        performance of location distinction for this use case.
\end{enumerate}

Under both use cases, in order to simulate MIMO antenna arrays of different
sizes and examine the associated performance of temporal signature-based
location distinction, we compile the MIMO link signatures, as in
(\ref{E:cplxMIMOtmpLinkSig}) and (\ref{E:MIMOtmpLinkSig}), from the subsets of
the SISO link signatures, CTLS and TLS, measured with $1 \times k$ and $k
\times k$ antenna arrays, where $k \in \{1,\ldots,8\}$. At the MIMO receiver,
channel measurements are made with a period $T_{r}$. For each measurement taken
at the receiver, we calculate the link signatures defined in Section
\ref{sub:link_signatures}. The number of channel measurements varies with the
receiver position.  For Experiment I, if there are $n+1$ measurements at a
given receiver location, the $n$th measurement is taken at $t = n T_{r}$, and
the $n$th link signatures are associated with this measurement time. In
Experiment II, $T_r$ varies slightly around a nominal value of 3.0 s, but we
have the exact position of the transmitter and receiver for each measurement.

\subsection{Experiment I} 
\label{sub:mimo_data}
The first experiment is conducted at Brigham Young University by Wallace et
al.\ \cite{wallace2006time}. MIMO channel data are collected using an 8x8 MIMO
channel sounder in which a multi-tone baseband signal is mixed with a carrier
frequency of 2.55 GHz and transmitted to stationary and moving receivers. The
transmitter is stationary for these measurements.  The multi-tone signal is
constructed as follows:
\begin{equation} \label{E:multiTone}
	x_{B}(t) = \sum_{i=0}^{B} \cos(2\pi f_{i}t+\theta_{i})
\end{equation}
where $B=39$ and
\begin{equation}
    f_{i}=(i+0.5) \mbox{ MHz}
\end{equation}
and $\theta_{i}$ is a fixed random phase shift between 0 and $\pi$ included for
each tone in order to spread the signal energy in time \cite{maharaj-cost}. The
signal $x_{B}(t)$ is multiplied by a Gaussian window to combat artifacts
generated by switching the signal on and off.

\begin{figure}[htbp]
\centering
\includegraphics[width=3.0in ]{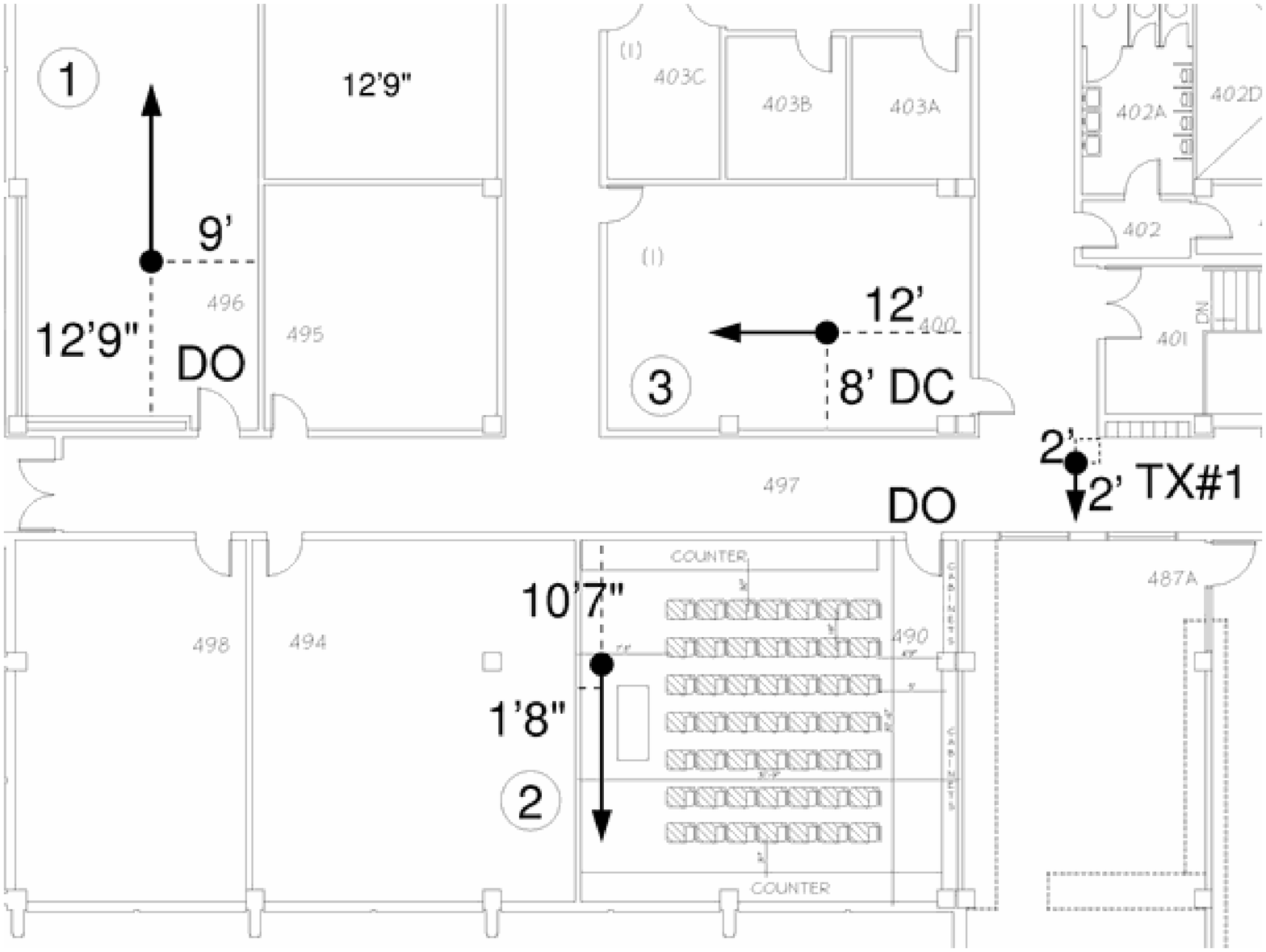}
\caption{Diagram of a subset of receiver locations from Experiment I. Circled
    numbers represent the receiver locations for individual measurement sets.
    DO or DC indicate door open or door closed, respectively.}
\label{F:floorplan}
\end{figure}

The transmitter and receiver each use a uniform circular array of eight
monopole antennas.  These arrays have a nominal element spacing of $\lambda/2$
(where $\lambda$ is the wavelength) and are well synchronized in both carrier
frequency and phase. The wideband channel frequency response $H(f)$ for each
antenna pair is computed by dividing the Fourier transform of the measured
signal by the Fourier transform of the known transmit signal and separating the
results into bins which correspond to the tones in the transmitted signal. The
wideband channel impulse response is calculated as 
\begin{equation}\label{E:ifftofHhat}
	h(n) = \mathcal{F}^{-1}\{H(f)\}. 
\end{equation}
where $\mathcal{F}^{-1}\{\cdot\}$ represents the inverse discrete-time Fourier
transform.  Channel measurements are collected at eight different receiver
locations on a single floor of an office building. Figure~\ref{F:floorplan} is
a diagram showing the first three receiver locations. The circled numbers
represent each location.

In this experiment the receiver is in motion while the transmitter is
stationary. We discussed in Section \ref{sec:introduction} applications which
detect a moving transmitter using stationary receivers, and the reciprocity of
the radio channel allows us to view these measurements as if this were the case
\cite{wilson2007channel,thiele1981antenna}. 

In the cases where the receiver is moving, it moves with a speed of 31.75
cm/sec. At each receiver location, between 390 and 585 measurements are made.
In the measurements made with a moving receiver, the multi-tone probe is sent
every 3.2 ms, or given the receiver speed of 31.75 cm/sec, every 1.016 mm.
These dense (spatially and temporally) measurements are the reason we delay
($D$) inserting the most recently measured link signature into the history
$\mathcal{H}$. As we show in Section \ref{sec:results}, the performance of
location distinction improves when this delay is increased, or equivalently,
when the current location of the receiver is further from its location during
the measurement of the most recent link signature in $\mathcal{H}$.


\subsection{Experiment II} 
\label{sub:Utah MIMO Data}
The second experiment is performed at the University of Utah. Channel
measurements are made using a MIMO-OFDM transceiver implemented with a National
Instruments vector signal generator (VSG) and vector signal analyzer (VSA) and
Labview software. 
    
The transmitted signal is designed to emulate the IEEE
802.11n standard \cite{80211n}.
It is an OFDM signal and has 64 subcarriers contained in a total bandwidth
of 20 MHz (312.5 kHz per subcarrier). These include four null subcarriers over
which the channel is not estimated (subcarrier indices -32, -31, 0, and 31).
Each data symbol is 4.0 $\mu$s long consisting of a 3.2 $\mu$s data symbol and
a 0.8 $\mu$s cyclic prefix. 

The frame (timing) synchronization, carrier offset
recovery, and channel estimation are aided by a preamble.
We use the greenfield preamble described in the physical layer specification of
the IEEE 802.11n standard, but we omit the high throughput signal field.  This
field is normally used to convey MAC information regarding the coding,
modulation scheme, etc., and isn't necessary for the channel estimation
required by this experiment. The preamble consists of an 8.0 $\mu$s periodic
signal with a short period (0.8 $\mu$s) for coarse carrier acquisition and
coarse frame synchronization. This is followed by 8.0 $\mu$s of a periodic
signal with a long period (3.2 $\mu$s) used for fine carrier acquisition and
fine frame synchronization.  Moose's method is used for frame synchronization
and carrier recovery \cite{MoosePaper,SDfSR}. 

The MIMO channel state is estimated using mutually orthogonal sequences. After
the long period signal, the transmitter sends to each antenna mutually
orthogonal sequences of symbols generated with Walsh-Hadamard codes for each
subcarrier. Each of these sequences has a duration of 4.0 $\mu$s, which
includes a 0.8 $\mu$s cyclic prefix.  A minimum mean-squared-error (MMSE)
channel estimation algorithm with a structure derived from the MMSE estimator
in \cite{KayBook} is employed.  Compared to the estimator in \cite{KayBook}, we
increase the number of transmit symbols used for estimating the channel from
two symbols (for a 2x2 system) to four symbols.

At the receiver, following carrier acquisition and frame synchronization, the
mutual orthogonality of the symbol sequences allows the receiver to quickly
invert the received signal information at each subcarrier by performing a
single matrix multiplication per subcarrier. This provides an estimate of the
channel response for each pair of antennas at each subcarrier, which are the
estimates used for this analysis.

\begin{figure}[htbp]
\centering
\includegraphics[width=2.5in]{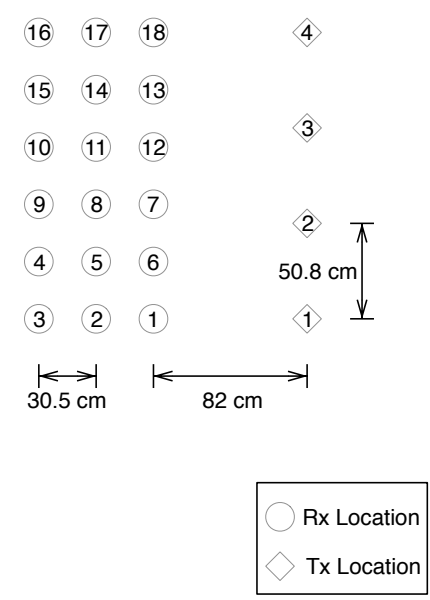}
\caption{Diagram of Experiment II. Circles represent receiver locations,
    diamonds represent transmitter locations. The outer line represents the
    wall of the room. Channel measurements are made at each
    transmitter/receiver location. Desks, equipment, and other scatterers are
    present, but not depicted in this diagram.}
\label{F:floorplanUtah}
\end{figure}

The data are collected in the Wireless Communication Lab at the University of
Utah, an open plan office lab containing desks, bookcases, chairs, and
measurement equipment. We take measurements at eighteen different receiver
locations and four different transmitter locations, as shown in
Figure~\ref{F:floorplanUtah}, resulting in a total of 3600 measurements of 72
distinct radio links. We choose a center frequency of 2.42 GHz and use whip
antennas separated by 15.24 cm for the transmitter and receiver antenna arrays,
placed at height of 0.91 m.


\section{\textbf{Results and Discussion}} 
\label{sec:results}
We present and discuss the results of Experiments I and II in the context of
four link signature characteristics. 

\subsection{Spatial Distance} 
\label{sub:physical}

\begin{figure}%
    \centerline{
            (a)
            \includegraphics[ ]{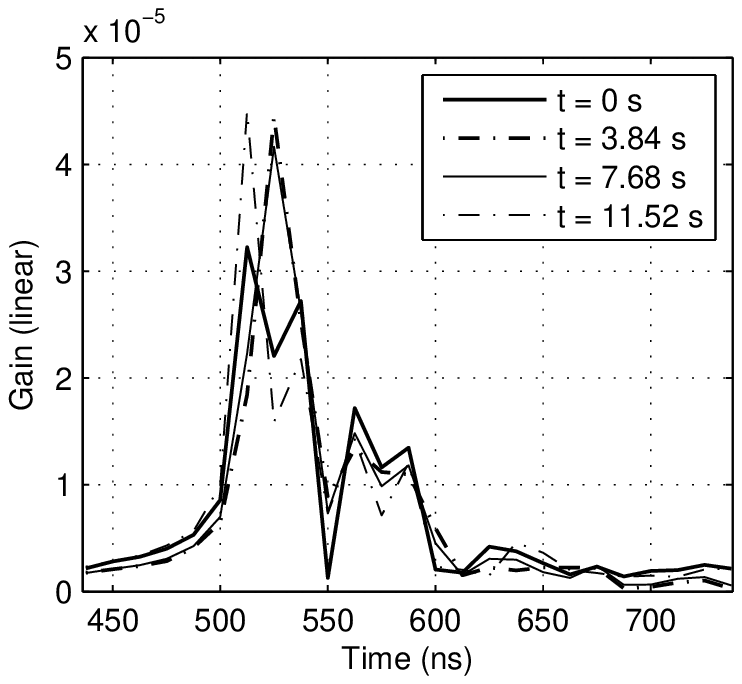}%
            (b)
            \includegraphics[ ]{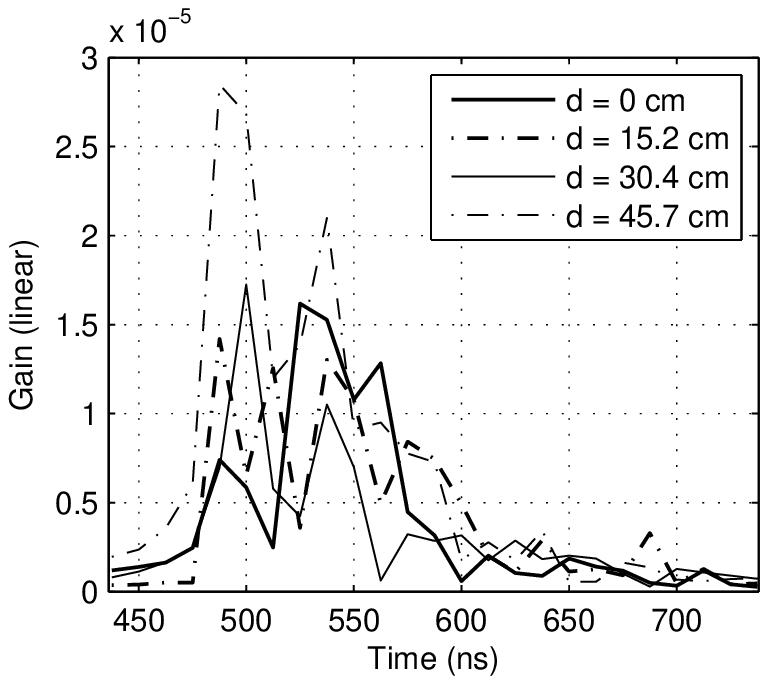}
        }
        \caption{Link signatures measured (a) over time at a stationary
            receiver and (b) at a moving receiver. The signatures measured at a
            moving receiver fluctuate more than those measured at the
            stationary receiver.}
\label{F:linkSigComp}%
\end{figure}

\begin{figure}%
    \centerline{
        (a)
        \includegraphics[ ]{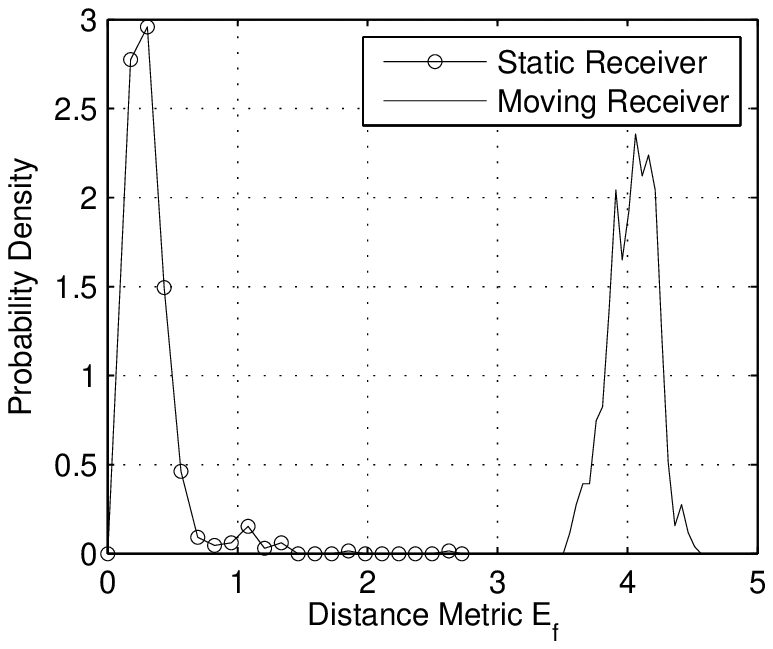}%
        (b)
        \includegraphics[ ]{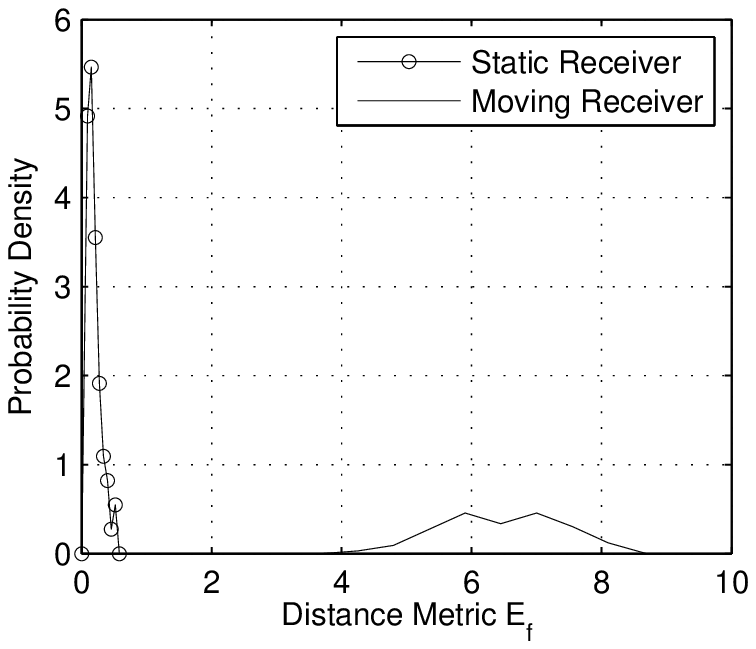}
    }
        \caption{Empirical distributions of $E_f$ for stationary and moving
            receiver from (a) Experiment I with 8x8 CTLS, and (b) Experiment II
            with the 2x2 CTLS. In both cases the mean difference metric for a
            moving receiver is significantly higher than for a stationary
            receiver.}
\label{F:distributions}%
\end{figure}

The results of both experiments show that differences in spatial location
between link signatures are more significant than the temporal variations in
link signatures measured for static receivers. In other words, changing the
position of the transmitter/receiver has a significant effect on the measured
link signatures.  Figure \ref{F:linkSigComp} shows the magnitudes of the 1x1
TLS measured at a stationary or moving receiver in Experiment I.  The variation
of the signatures for the moving receiver is more significant.  In the case of
the MIMO TLS, the same effect can be seen in the empirical distributions of the
difference metric (\ref{E:distMetric}). These distributions are shown in
Figure~\ref{F:distributions}(a). The mean difference metric is much higher in
the case of a moving receiver. The same result can be seen in the empirical
distributions of the difference metrics calculated for Experiment II. These
distributions are shown in Figure \ref{F:distributions}(b). 

\begin{figure}[]
    \centerline{
        (a)
        \includegraphics[ ]{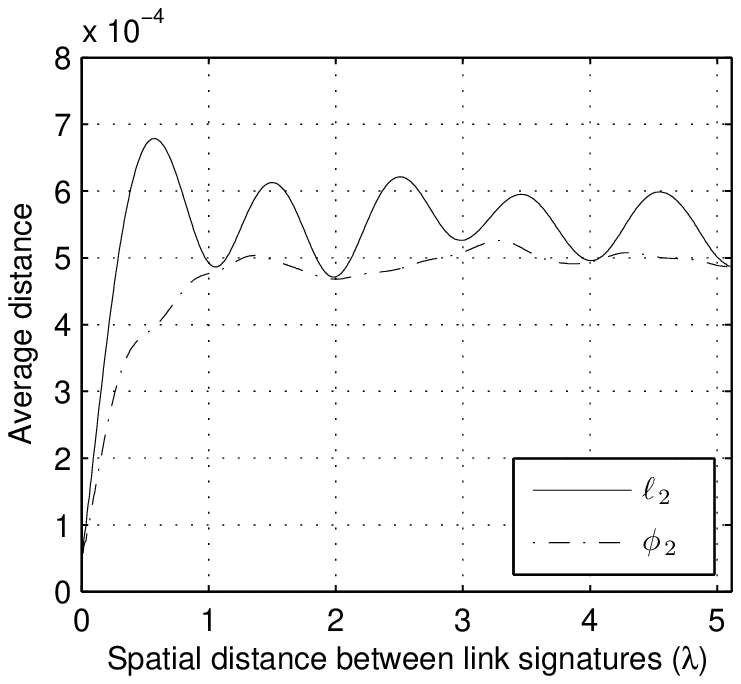}
        (b)
        \includegraphics[ ]{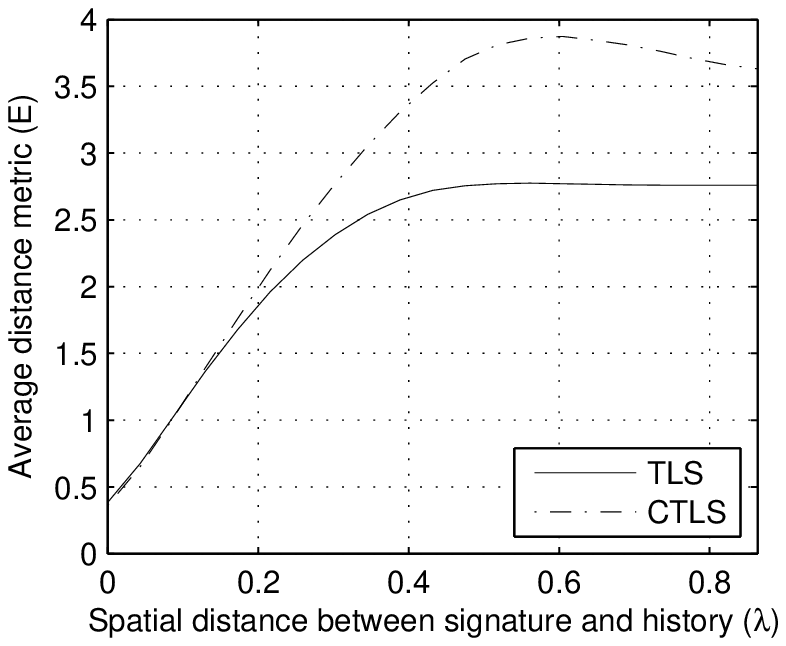}
    }
    \caption{(a) Average $\ell_2$ and $\phi_2$-distances between 8x8 MIMO CTLS
        as a function of spacial separation. The average $\ell_2$-distance
        peaks at a receiver separation of roughly $\lambda/2$. (b) Average
        difference metrics $E$ for 8x8 CTLS/TLS as a function of spatial
        separation.}
    \label{F:avg_distances}
\end{figure}

Figure~\ref{F:avg_distances}(a) shows the average $\ell_2$ and $\phi_2$
distances between 8x8 MIMO CTLSs as a function of receiver separation where the
$\phi_2$ distance is defined in (\ref{E:phi2}).  The average $\ell_2$-distance
reaches a maximum at a separation of approximately $\lambda/2$, and then
oscillates with a period of $\lambda$. This result agrees with a result of the
Clarke fading model, which assumes incoming multipath are uniformly distributed
about the receiver \cite{rappaport1996wireless}.  The average $\phi_2$-distance
peaks at a receiver separation of about $\lambda$ and the oscillation is
mitigated by the phase rotation inherent in the $\phi_2$-distance.
Figure~\ref{F:avg_distances}(b) shows the average difference metrics $E$ as a
function of receiver separation. These results indicate that the difference
metrics perform best in the case where the receiver has moved a wavelength or
more between measurements. In the context of an impersonation
attack, this is typically the case. 

\subsection{History Size} 
\label{sub:history}
The size of the history buffer $\mathcal{H}$ is a parameter which should be
chosen in order to provide the best location distinction performance. For this
work, we select a range of history sizes to examine in both experiments and
identify the best size heuristically. However, the optimal number of signatures
to inlcude in the history is a function of the the difference metric being used
and the distribution of the differences measured under $\mathbb{H}_0$ and
$\mathbb{H}_1$. Because of the minimum operator in (\ref{E:distMetric}),
increasing the history size can only lower the average difference metric under
both hypotheses.

\begin{figure}[htbp]
    \centerline{
        (a)
        \includegraphics[ ]{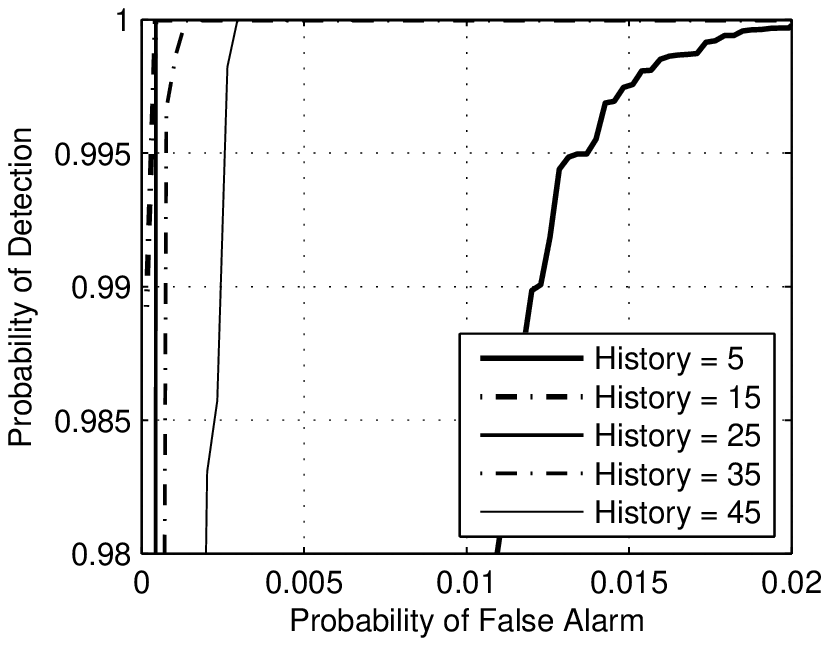}
        (b)
        \includegraphics[ ]{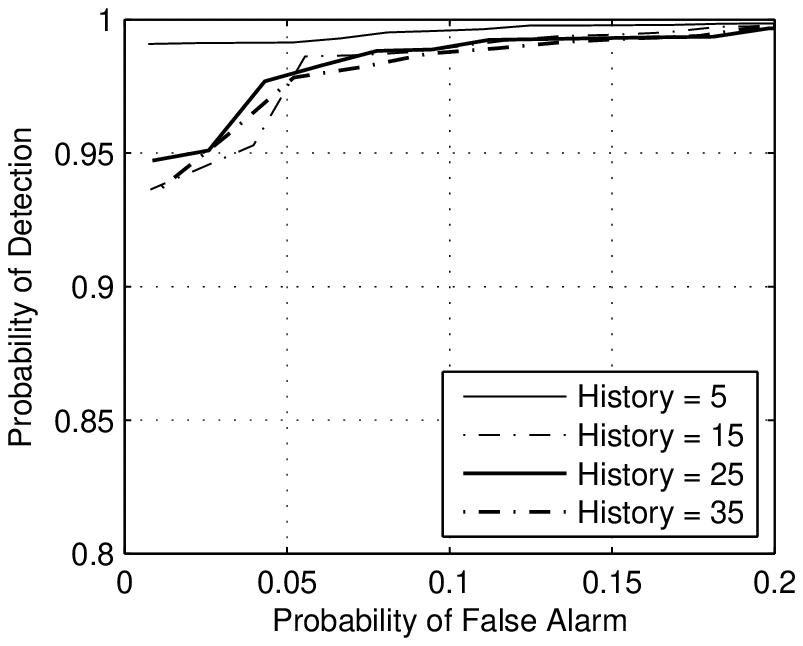}
    }
    \caption{ROC curves for (a) Experiment I: 8x8 MIMO CTLS and (b) Experiment
        II: 1x1 CTLS for various history sizes. In Experiment I, a history size of 15 link
        signatures yields the best performance. In Experiment II, a history size of 5 link
        signatures yields the best performance.}
    \label{F:histories}
\end{figure}

Figure~\ref{F:histories}(a) shows the receiver operating characteristic (ROC)
curve of the location distinction algorithm for the 8x8 MIMO CTLS and various
history sizes. In this case, the best performance corresponds to a history
containing fifteen previous link signatures. Figure \ref{F:histories}(b) shows
the ROC curve of the location distinction algorithm for the 2x2 CTLS of
Experiment II and various history sizes. In this case, a history size of five
offers the best performance. 

The differences measured under $\mathbb{H}_0$ in Experiment I have a
significantly higher mean/variance than those measured under the same
hypothesis in Experiment II, indicating that the temporal variations of the
link signatures measured for a stationary receiver in Experiment I are more
prominent than those in Experiment II. Therefore, a larger history size is
necessary in Experiment I in order to capture the temporal variations of the
stationary receiver.

\subsection{Number of Antennas} 
\label{sub:num_antennas}
The results show that as the size of the MIMO antenna array is increased, the
performance of the location distinction algorithm improves. This is consistent
with the simulation results of \cite{xiao2008mimo}, which used ray-tracing
simulations to show that the average miss rate in a location distinction system
decreases with the number of antenna elements.

\begin{figure}[htbp]
    \centerline{
        (a)
        \includegraphics[]{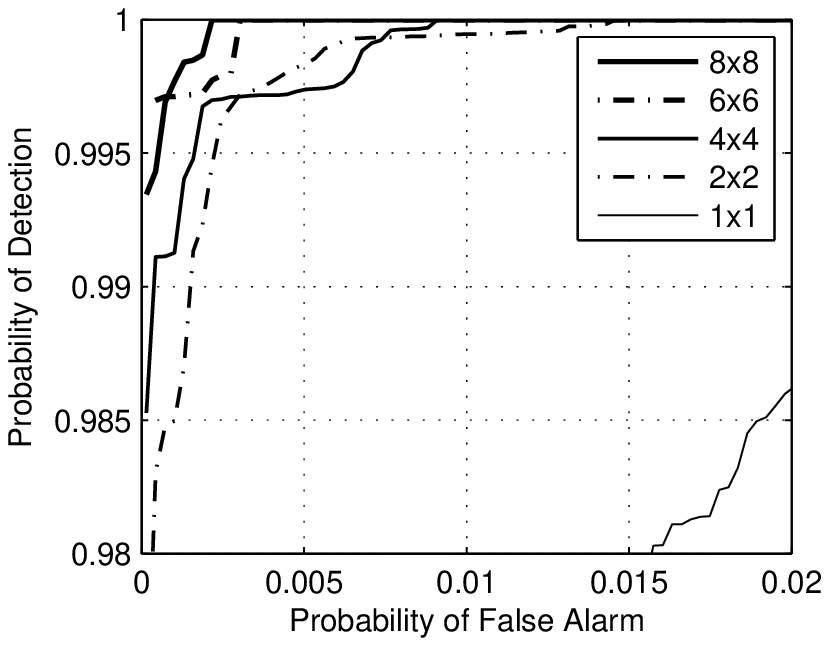}
        (b)
        \includegraphics[]{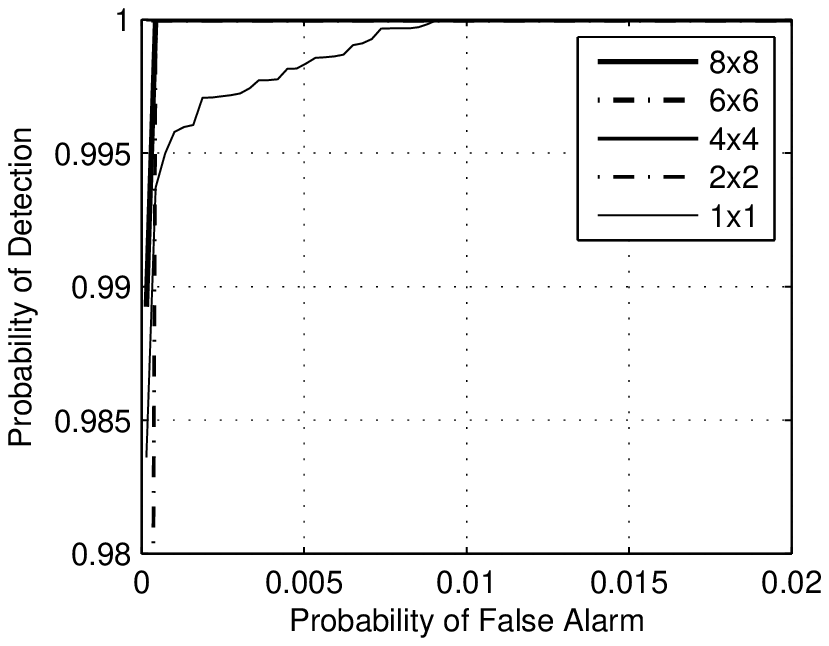} 
    }
    \centerline{
        (c)
        \includegraphics[]{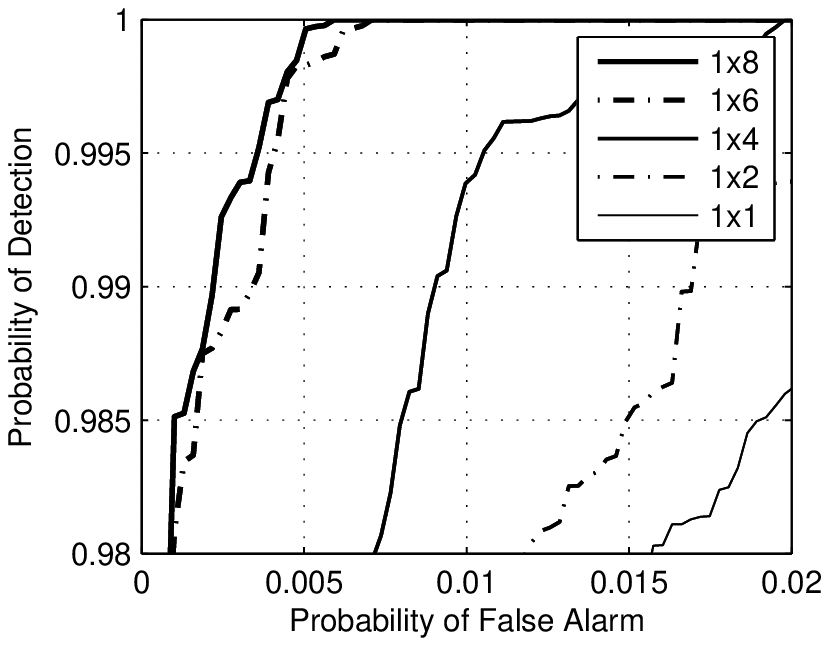}
        (d)
        \includegraphics[]{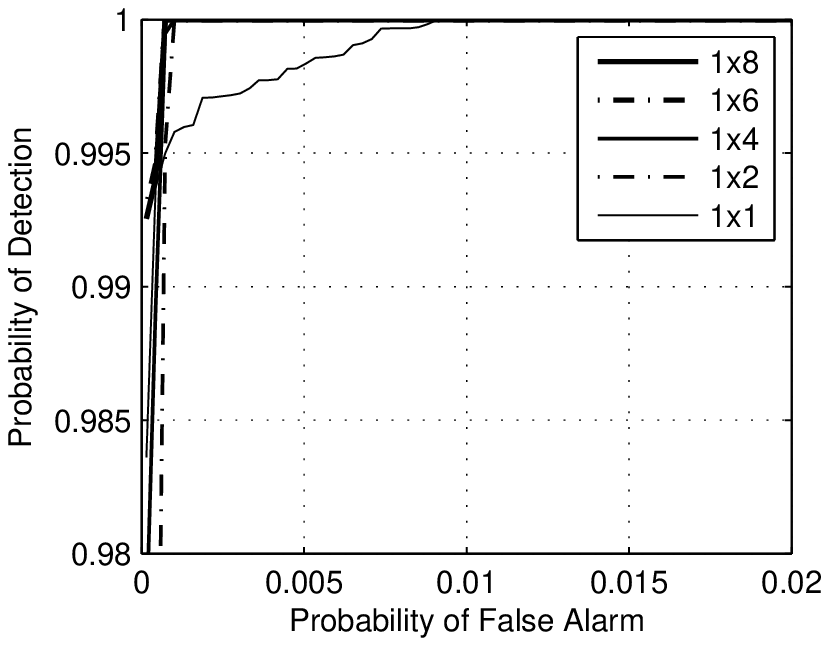}
    }
    \caption{ROC curves for (a) MIMO TLS (b) MIMO CTLS (c) SIMO TLS and (d) SIMO CTLS for various antenna
        array sizes. Location distinction performance improves with the number
        of antennas and the MIMO CTLS performs better than the MIMO TLS. The
        SIMO signatures nearly match the performance of the MIMO signatures.}
\label{F:pkVsCpk}
\end{figure}

\begin{figure}[htbp]
\centering
\includegraphics[ ]{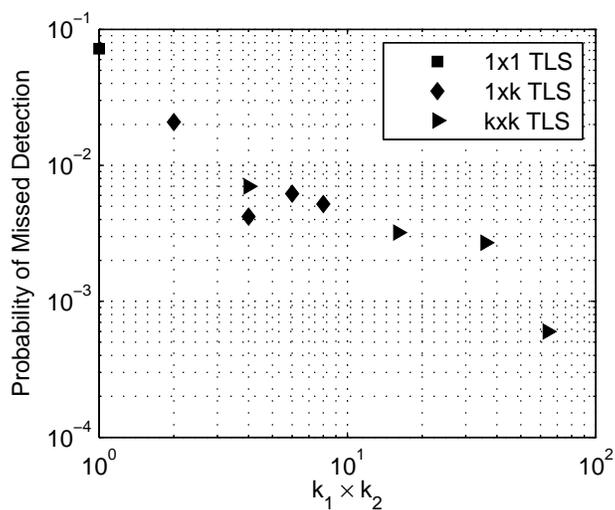}
\caption{Experiment I: Probability of missed detection for a $2\times 10^{-3}$
    probability of false alarm vs. $k_1 k_2$ for different SISO, MIMO,
    and SIMO arrays.}
\label{F:pmvsnumch}
\end{figure}

Figures \ref{F:pkVsCpk}(a) and \ref{F:pkVsCpk}(b) show the location distinction
ROC curves for the data from Experiment I and different sized MIMO antenna
arrays. Figures \ref{F:pkVsCpk}(c) and \ref{F:pkVsCpk}(d) show the ROC curves
for the same experiment, but using SIMO arrangements. The trend in these
figures is toward better location distinction performance with the increase in
size of the MIMO antenna array. Figure \ref{F:pmvsnumch} shows miss rates for a
given false alarm rate and various SISO, SIMO, and MIMO arrays. The miss rates
appear to follow the inverse power law 
\begin{equation}
    P_M = \frac{b}{(k_1 k_2)^m}
    \label{E:pmrule}
\end{equation}
where $b$ and $m$ are parameters that define the rate that the probability of
missed detection approaches zero with the number of MIMO channels. A
least-squares approximation yields $b \approx 10^{-1.44}$ and $m \approx 0.93$
for the data in Figure \ref{F:pmvsnumch}. As a rule of thumb, the achievable
miss rate for a constant false alarm rate is approximately inversely
proportional to $k_1 k_2$, the number of channels.

\begin{table}[htbp] 
    \caption{$P_{M}$ for $P_{FA}=10^{-2}$ for Experiments I and II.} 
    \centering      
    \begin{tabular}{c | c c c | c c c}
        \hline\hline                        
        & \multicolumn{3}{c}{Experiment I} & \multicolumn{3}{c}{Experiment II}\\
        \hline  
        $(k_1,k_2)$ & MIMO TLS & MIMO CTLS & CTLS/TLS & MIMO TLS & MIMO CTLS & CTLS/TLS \\
         & $P_{M}$ & $P_{M}$ & Improvement & $P_{M}$ & $P_{M}$ & Improvement\\ [0.5ex] 
        \hline                   
        (1,1) & 0.032 & $\leq$ 0.00024 & $\geq$ 133x & 0.0323 & 0.0092 & $\approx$ 3.5x \\   
        (2,2) & 0.0005 & $\leq$ 0.00024 & $\geq$ 2x & $\leq$ 0.0003 & $\leq$ 0.0003 & N/A \\[1ex]    
        \hline\hline
        MIMO Improvement & $\approx$ 64x & N/A & & $\geq$ 108x & $\geq$ 31x \\
        \hline\hline      
    \end{tabular} 
\label{T:performance}   
\end{table}

The most drastic improvement in the miss rate occurs in the change from a SISO
channel to a 2x2 MIMO or 1x4 SIMO channel. Table~\ref{T:performance} shows the
improvement of the location distinction algorithm in a 2x2 MIMO channel over
the SISO channel in Experiment I.  There is as much as a 108-fold reduction in
the miss rate for a constant false alarm rate when changing from SISO to 2x2
MIMO.


\subsection{MIMO CTLS and TLS} 
\label{sub:ctlstls}
In comparing Figures \ref{F:pkVsCpk}(a) and \ref{F:pkVsCpk}(b), it is also apparent
that the MIMO CTLS and its associated difference metric leads to better
performance than the MIMO TLS in Experiment I. Table~\ref{T:performance} shows
the improvement of the location distinction algorithm when using the MIMO CTLS
instead of the MIMO TLS. Using the MIMO CTLS results in as much as a 133-fold
reduction in miss rate for a constant false alarm rate. 

This result is also confirmed in Experiment II, as shown in
Table~\ref{T:performance}. In Experiment II, the 1x1 CTLS results in a
3.5-fold improvement in miss rate over the 1x1 TLS. The 2x2 TLS and 2x2 CTLS
both reach the lowest measurable miss rate in Experiment II.

\subsection{Link Signature Bandwidth} 
\label{sub:Link Signature Bandwidth}
Another crucial parameter in both experiments, and typically a limiting factor
in radio design, is system bandwidth. We examined the performance of the
location distinction algorithm over a range of bandwidths by varying the number
of tones included in the IFFT of the frequency-domain measurements from
Experiment I.  This is similar to varying $B$ in (\ref{E:multiTone}). 

\begin{figure}[htbp]
\centering
\includegraphics[ ]{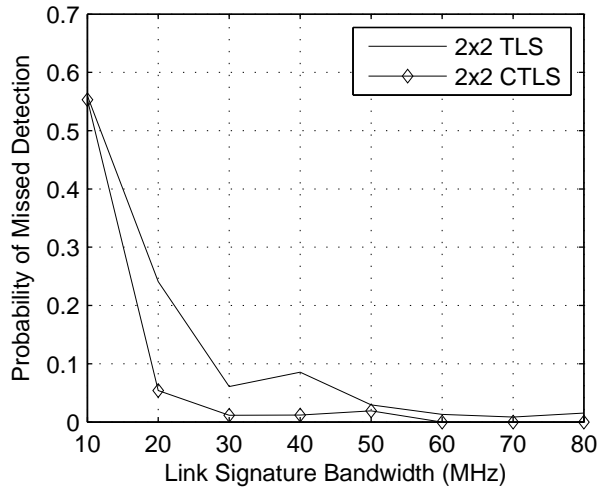}
\caption{Location distinction miss rate vs. link signature bandwidth for a
    $7\times 10^{-4}$ false alarm rate in Experiment I. Increasing bandwidth offers diminishing returns.}
\label{F:pmvsbw}
\end{figure}

Figure~\ref{F:pmvsbw} shows that performance typically improves with bandwidth,
but it does so with diminishing returns. This is consistent with the simulation
results of \cite{xiao2008mimo}, which show that the miss rate of a location
distinction system decreases with system bandwidth, but that the performance
gain of MIMO over SISO also decreases, because at high bandwidths the SISO link
signatures offer more decorrelation.  

\begin{figure}[htbp]
\centering
\includegraphics[ ]{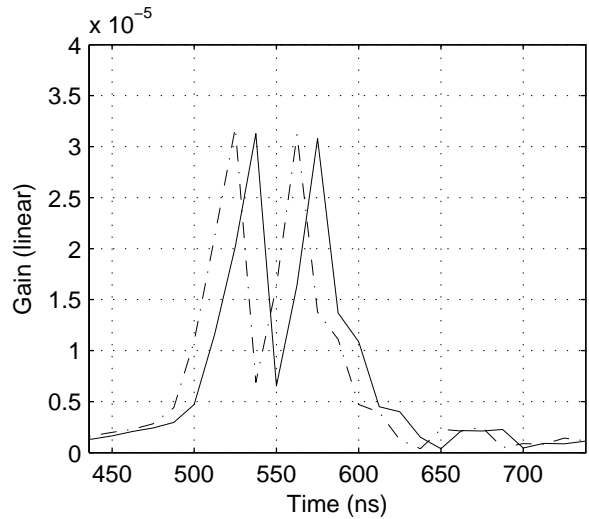}
\caption{Two consecutive link signatures with 80 MHz bandwidth showing the
    results of a timing-synchronization error. The time-resolution of
    high-bandwidth link signatures cause an increased impact on location
    distinction performance.}
\label{F:jitter}
\end{figure}

However, at high bandwidths the algorithm is more sensitive to
timing-synchronization errors that might be hidden by lower bandwidth
signatures. Figure~\ref{F:jitter} shows an example of two consecutively
measured link signatures that exhibit this effect. These errors cause small
drops in performance.  The higher bandwidth of the link signatures measured in
Experiment I (80 MHz) allows for better location distinction performance, but
the results for the 2x2 MIMO link signatures of Experiment II (20 MHz) still
offer a $3\times 10^{-4}$ probability of missed detection for a $7\times
10^{-3}$ probability of false alarm.


\section{\textbf{Related Work}} 
\label{sec:related_work}
The papers discussed in this section have contributed to this work in different
aspects. The most closely related work is that of Patwari et al.\
\cite{patwari2007robust} and Zhang et al.\ \cite{zhang2008advancing}. In these
two papers, a temporal link signature is defined to be used in the context of
multiple transmitters/receivers and then refined to include phase information.
We compliment that work by showing that a single MIMO transmitter/receiver pair
can be used to perform reliable location distinction, and that lower false
alarm rates are possible using a single receiver, when the communication system
is a 1x2 or 2x2 MIMO system. In \cite{zhang2008advancing}, the authors report a
$9\times 10^{-3}$ miss rate for a 0.01 false alarm rate using three receivers.
For the same false alarm rate, we are able to achieve a $3\times 10^{-4}$ miss
rate using a single receiver and the 2x2 MIMO CTLS with less bandwidth.  This
net reduction in system complexity may enable location distinction in future
wireless networking systems.

In \cite{zhang2008advancing} a \emph{complex temporal link signature} is
defined which allows for the exploitation of the phase information in the
CIR. However, not all of the phase information represented by the link
signature is due to the channel. Some phase shifts occur due to a lack of time
and/or frequency synchronization between the transmitter and receiver. The
distance between two link signatures which minimizes the contribution of random
phase shifts corresponds is shown to be (\ref{E:phi2}). Zhang et al.\ call this the
$\phi_2$-distance. It is not necessary to apply (\ref{E:phi2}) to the data
gathered in Experiment I, because it is phase-synchronous, but we do apply it
to the data from Experiment II.

In \cite{xiao2008mimo}, Xiao et al.\ present ray-tracing simulation results for
MIMO location distinction in defense of impersonation attacks in an office
building. They assume that channel measurements made in the frequency domain
are distributed as complex Gaussian random variables and derive ideal change
metrics based on this assumption. We extend this work by offering an
experimental validation of MIMO location distinction using two MIMO testbeds. 

In \cite{li2006securing}, Li et al.\ propose some of the underlying ideas of
this work, namely, that characteristics of the radio channel (rapid
de-correlation in space, time, and frequency) can be exploited to secure
wireless networks. They offer methods of probing the channel in order to
determine, based on the channel gains between transmitters and receivers,
whether or not communications are coming from an authentic user or a would-be
attacker. Using the USRP/GNU Radio and a simple change-point detector, they
show that they are able to detect a change in the wireless link via channel
gains and thereby detect a possible spoofing attack.

In \cite{faria2006detecting}, Faria and Cheriton utilize similar principles in
designing a method for identifying a transmitter by its \emph{signalprint},
which consists of a vector of RSS values. These RSS values are gathered using
wireless access points as sensors and a central authentication server for
cataloging and comparing signalprints. Their results show that a stationary
transmitter will produce a consistent signalprint and thereby allow for
discrimination between authentic users and attackers whose signalprints will
vary significantly because they are located in a different position in the
multi-path fading channel. The signalprint is limited in that it may be unable
to detect attackers located near authentic transmitters, because they may have a
similar signalprints. 


\section{\textbf{Conclusion and Future Work}} 
\label{sec:conclusion}
In this paper we show that techniques for location distinction can be applied
to a MIMO channel. Using two distinct measurement sets, we show that a simple
linearization of the link signatures for each transmitter/receiver antenna pair
can be used to form MIMO link signatures, and that difference metrics can be
used to determine whether or not a wireless device has changed position. Our
results show that the presented MIMO location distinction framework can be used
to discern a stationary transmitter from a moving transmitter with accuracy
better than any previously reported experimental results. We also show how the
adjustment of the parameters of the location distinction algorithm (history
size, spatial separation, complex/magnitude-only signatures, and number of
antennas) affect the performance of the algorithm. 

In addition to the promising results we have shown, it will be beneficial to
further characterize the link signatures used for location distinction and
explore other difference metrics. For instance, our current difference metric
uses the minimum Euclidean or $\phi_2$-distance between the most recent link
signature and those in the history $\mathcal{H}$. This tends to increase the
miss rate in the context of noisy measurements. A weighted average of
distances, such as the Mahalanobis distance may offer better performance. A
broader experimental analysis of link signatures and their temporal and spatial
variations will facilitate the design of better difference metrics.

\bibliographystyle{IEEEtran}
\bibliography{main.bib}  
\end{document}